Plant evolution on rock outcrops and cliffs: contrasting patterns of diversification following edaphic specialization

Isaac Lichter Marck[1,2,*]

[1]Department of Ecology and Evolutionary Biology, University of California, Los Angeles

[2]Jepson Herbarium & Department of Integrative Biology, University of California, Berkeley

*Corresponding author: ilichtermarck@gmail.com

Running head: Plant evolution on cliffs and rock outcrops

**Abstract**

Sheer cliffs on mountains and in deep canyons are among the world's most iconic landmarks but our understanding of the enigmatic flora that lives in these vertical-rock landscapes remains fragmented. In this article, I review and synthesize recent studies on the evolution of specialization onto bare rock and its consequences for plant diversification. Putative adaptations commonly associated with growth on bare rock include specialized root structures, stress tolerant leaf traits, and reduced dispersibility. Fitness trade-offs are a principal explanation for edaphic specialization, but adaptation to bare environments stands apart as a precursor environment to specialization in other stressful habitats, such as chemically harsh soils. In species-level phylogenies, many rock specialist plants are evolutionarily isolated and form the sister lineage to congeners found on decomposed substrates, suggesting that specialization onto bare rock may be an evolutionary trap or dead-end. In other cases, rock specialists form diverse clades, suggesting archipelago speciation or ecological release enabled by innovations in stress tolerance. Today, in environments severely impacted by poor management and introduced organisms, such as islands, cliffs serve as important refuges for stress tolerant endangered plants. New technology for climbing safety and drone assisted reconnaissance open new possibilities for the discovery and conservation of cliff plants in these landscapes but come with the inherent risks of increasing their accessibility. The future success of efforts to save plants from extinction may depend on our understanding of the unique and resilient flora endemic to cliffy places.

**Key words:** biogeography; chasmophyte; conservation; climbing; drones; inselbergs; macroevolution; OCBILs; saxicolous plants; rupicolous plants.

**INTRODUCTION**

The dispersed information from floras, checklists, and ecological studies (reviewed by Larson et al., 2000) suggests that cliffs contribute more to the biodiversity of a region than their surface coverage would indicate and are a promoter of heightened endemism in biodiversity hotspots (Buira et al., 2021; Mota et al., 2020; Silveira et al., 2016; Vasconcelos et al., 2020; Lambers, 2014; Baldwin, 2014; Baldwin et al., 2017; Fitzsimons and Michael, 2017; Allsopp et al., 2014; Fig.1). The realization that rock outcrops harbor a diverse, unique, and underappreciated flora has stimulated a flurry of interest in their unique biodiversity (Hopper, 2009; Mucina and Wardel-Johnson, 2011). This has led to recent efforts to review key aspects of their biology, including ecology (Porembski and Barthlott, 2000), patterns in biodiversity (Silveira et al., 2016), conservation (Fitzsimons and Michael, 2016), and vulnerability to climate change (Corlett and Tomlinson, 2020). One area that has received less attention, however, is the evolution of edaphic endemism onto bare rock and its contribution to diversification (Corlett and Tomlinson, 2020). What are the forces that have generated such a diversity of plants that grow on bare, rocky outcrops? Here, I review and synthesize recent research concerning the evolution of rock specialists to learn about 1) the origins of specialization onto bare rock, 2) consequences of this specialization on lineage diversification, and 3) implications for conservation.

Exploration and documentation of plants in difficult to access cliffs and deep canyons is a perennial source for discoveries of new species to western science (e.g. Siniscalchi et al., 2018; Huang et al., 2019; Fig. 2). Many areas with abundant cliffs remain botanical black holes with few documented collections due to their rough, inaccessible terrain. In addition, the recent explosion of studies that use spatial data tend to neglect cliffs and most rocky outcrops because vertical environments are not apparent on 2 dimensional maps and edaphic factors often neglected in climatic niche models (de Casas et al, 2016; Corlett and Tomlinson, 2020; Folk et al., 2021). A more general difficulty in the subfield of cliff plants has been the proliferation of different terms for rock outcrops and rock specialists in different places, which fractures the literature and hampers progress towards a clear framework for hypothesis testing (Mucina and Wardel-Johnson, 2011; Table 1).

    I surveyed the literature for empirical studies that have focused on rock specialization, evolution of putative adaptations to rock, or generated molecular phylogenies that include rock-dwelling plants by searching the ISI Web of Science database and google scholar using the various technical terms for plants on bare, rocky outcrops (see Table 1). I also reviewed previous synthesis papers that deal with plant evolution or ecology on different harsh edaphic substrates, including serpentine (Kay et al., 2011), gypsum (Moore et al., 2014), white sands (Fine and Baraloto, 2016), and limestone (Kruckeberg, 2004). In this synthesis, rock outcrops refer to patches of unweathered bedrock that emerge from the soil surface and cliffs refer to rock outcrops with a vertical or overhanging exposure. I adopt the terms cliff specialist, rock specialist, and rock-dwelling to refer to plant taxa that are found always ('specialist') or mostly ('dwelling') on rock outcrops. For simplicity, I focus on inland cliffs in this review, rather than maritime cliffs or the man-made outcrops caused by road cuts and development. Finally, I draw mostly from information on flowering plants based on my background and not because Angiosperms are the sole inhabitants of rock outcrops; in fact,

Gymnosperms and cryptogams are extremely common on rock outcrops. Lichens, for example, have been the focus of some of the most intensive ecological research on cliff ecology in recent years (e.g. Büdel and Friedl, 2021; Henrie et al., 2022; Rutherford and Robertus, 2022).

**PUTATIVE ADAPTATIONS FOR PLANT GROWTH ON BARE ROCK**

Growth on bare rock requires that plants overcome multiple non-mutually exclusive abiotic challenges, including increased exposure to wind, vertical orientation, increased UV radiation, increased aridity, decreased soil moisture, and lack of available soil and macronutrients (Larson et al., 2000; Kruckeberg, 2004; Cacho and Strauss, 2014; Bátori et al., 2019; Corlett and Tomlinson, 2020). Biotic interactions may also play a role in making this a challenging environment for plant growth, with increases in herbivory due to apparency, reduced pollination due to decreased pollinator abundance, lack of mycorrhizae, and decreased abundance of facilitative nurse plants (Strauss and Cacho, 2013; Cacho and Strauss, 2014; Lewis and Schupp, 2014; Cacho and MacIntyre, 2020; Abrahão et al., 2020). Putative adaptations for growth on bare rock include stress tolerant leaf traits, such as dense hairs, small waxy leaves, or lack of leaves, a ceaspitose growth form, shifts towards self-compatibility, and heightened anti-herbivore defenses (Axelrod, 1972; Grime, 1977; Baskin and Baskin, 1988; Cacho and Strauss, 2014; Kay et al., 2017; De Paula et al., 2019; Lichter-Marck et al., 2020; Vargara-Gomez et al., 2020; Table 2).

Like organisms restricted to islands, many lineages of rock dwelling plants appear to be under selection for reduced dispersibility (Hopper et al., 2016; Burns, 2020), including loss of dispersal elements leaving dry fruits without adaptations for anemochory or zoochory (Silviera et al., 2020; Lichter-Marck et al., 2020). Limited dispersal is reflected in high fixation rates (Fst) across 80 population genetic studies from species found in Campos Rupestres in Brazil (Silveira et al., 2020) and similar patterns have been found in rock specialists elsewhere (e.g. Folk and Freudenstein, 2015). In an extreme example of reduced dispersibility, several disparate lineages of rock specialist plants in western North America have evolved a behavior in which the pedicel bends back as fruits mature, self-sowing in the rock face behind the parental plant (Rebman et al., 2016; Fig.3). Such a clear example of reduced dispersibility could be attributed to low survival off of rock outcrops (e.g. Schenk, 2013), but rock outcrops also offer few opportunities for recruitment, which might encourage strategies for reseeding around the parent plant (Larson et al., 2000).

Another similarity with island organisms is the prevalence of woody tissues among rock specialists (Burns, 2022), which may reflect the need for specialized root structures for anchoring and nutrient uptake in unweathered substrates (Abrahão et al., 2014; Hopper et al., 2016; Abrahão et al., 2020; Shi et al., 2020). A woody caudex is a common feature in rock specialists and may serve a dual function for anchoring in cracks as well as energy storage during inclement seasons or fires (Poot et al, 2012). In the rock daisies (Perityleae B.G. Baldwin), a north American desert tribe in the sunflower family (Compositae) with numerous rock specialists, the woody caudex is a complex morphological trait whose history is evolutionarily correlated with transitions to rock specialization, suggesting a key role for this structure in facilitating growth on bare rock (Lichter-Marck and Baldwin, 2022b). Roots of the tropical, and often rock dwelling family Velloziaceae J. Agardh. are highly specialized structures capable of carboxylate release to actively dissolve quartzite rocks and release phosphorus in the process (Teodoro et al., 2019) and these structures are especially apparent in taxa found on rock (Abrahão et al., 2020). Active nutrient uptake from unweathered bedrock is an active subject of research and may be a more widespread strategy among rock specialists, and plants in general, than previously believed (Lambers et al., 2012; Dawson et al., 2020).

**EDAPHIC SPECIALIZATION ONTO BARE ROCK**

How did such a variety of putative adaptations to cope with growth on bare rock evolve? Following two decades of discussion about the mechanisms of ecological specialization, fitness tradeoffs have emerged as a prime explanation (Kursar and Coley 2003, Coley et al., 2005; Fine et al., 2006; Cacho and Strauss, 2014). Tradeoffs occur when adaptation to one environment incurs costs to growth in another (Box 1). In the case of rock outcrops, plants that invest heavily in traits conferring survival on bare rock are expected to thrive on these stressful habitats but face energetic costs that will cause them to grow slowly and be excluded by more vigorous plants on surrounding decomposed substrates. Indirect evidence for the importance of trade-offs for edaphic specialization onto rock outcrops includes the fact that rock dwelling plants grow extremely slowly and reach surprisingly old ages (Larson et al. 1999), and that many possess specialized root morphology that could disadvantage them in other habitat types (Poot et al., 2012; De Paula et al., 2015).

Fitness tradeoffs in endurance and growth for plants that grow on rock outcrops have rarely been rigorously tested, however, recent research on the jewelflowers (*Streptanthus*) has included studies on bare habitats, a potential precursor to establishment on serpentine soils (Cacho & Strauss, 2014; Sianta and Kay, 2019). Bare habitats are a broader category than bare, rocky outcrops and refer to areas with a high degree of unvegetated space due to various sources of habitat stress, including soil chemistry, soil texture, climate, and natural enemies, but in many cases, it is closely associated with habitat rockiness (Cacho and Strauss, 2014; Sianta and Kay, 2019; Cacho and MacIntyre, 2020). In their recent analysis of *Streptanthus,* Cacho and Strauss (2014) found evidence for a correlation between slow growth and microhabitat bareness that evolved prior to colonization of chemically harsh serpentine soils. In a broader study, Sianta and Kay (2019) similarly found that serpentine specialists occupied more bare soil than their serpentine tolerator relatives. These studies provide support for a provocative hypothesis: that occupation of physically challenging environments is a prerequisite to colonization of harsh chemical substrates in plants.

**CONTRASTING PATTERNS OF DIVERSIFICATION FOLLOWING EDAPHIC SPECIALIZATION**

The idea that specialization onto bare micro-environments may have pre-adapted lineages for other harsh environments is not a new one. Over 50 years ago for example, Daniel Axelrod (1972) described a visit to Brazil's Atlantic rainforest, where he observed drought resistance plants characteristic of deserts, such as *Acacia* (Fabaceae) and *Opuntia* (Cactaceae), growing on bare sandstone outcrops in a matrix of tropical forest.

Based on these observations, Axelrod (1967, 1972) proposed that the North American desert flora was derived from drought tolerant plants that originally evolved on edaphically dry microsites in more densely vegetated biomes. The molecular revolution has provided us with a wealth of information with which to evaluate hypotheses like Axelrod's about the effects of rock-specialization on a lineage's evolutionary trajectory.

A literature scan in search of rock specialist plant taxa included in species-level molecular phylogenies shows that lineages conform to two predominant patterns (Table 3). On the one hand are clades in which rock specialist plants are evolutionarily isolated and more closely related to species found on decomposed soils than they are to other rock specialists. For example, among monkeyflowers (Phrymaceae Schauer) are rock specialists that evolved independently in the western great basin (*Diplacus rupicola* (Coville & A.L.Grant) G.L.Nesom & N.S.Fraga; Fig.2a) and Colorado plateau (*Erythranthe eastwoodiae* (Rydb.) G.L.Nesom & N.S.Fraga) that are more closely related to congeners found in nearby decomposed soils (Beardsley et al., 2004). On the other hand, are genera with many closely related species entirely restricted to bare, rocky environments. Exemplary cases of extensive diversification on rock outcrops include the stonecrops (Aizoaceae Martinov) of South Africa (Ellis et al., 2006), the Pitcarnioid Bromeliads (Bromeliaceae Andreanae) of South America (Palma-Silva et al., 2011; de Paula et al., 2016; Givnish et al., 2016; de Casa et al., 2016), rock daisies (Perityleae) of western North America (Powell, 1973; Lichter-Marck and Baldwin, 2022b; Fig. 4), and saxifrages (Saxifragaceae Juss.) of Holarctic temperate environments (de Casas et al., 2016; Folk et al., 2021).

Illustrations of contrasting patterns of diversification following specialization onto bare rock are illustrated in Box 2 along with mechanistic hypotheses and specific predictions that could be tested with macroevolutionary studies. These contrasting patterns match patterns of plant endemism on white sands soils in the Amazon basin (Fine and Baraloto, 2016) and suggest that edaphic specialization can have very different consequences in different lineages. The first case, where rock specialists are phylogenetically isolated and closely related to congeners found on decomposed substrates, is consistent with the hypothesis of evolutionary traps (Snogerup, 1971). This hypothesis states that edaphic specialization onto rock outcrops evolves in one, irreversible direction, because adaptation to bare rock requires complex phenotypic changes that would be difficult to undo. This first pattern may also be consistent with the evolutionary dead-ends hypothesis (Vamosi et al., 2014), under which rock-specialization is seen as a risky strategy on long time scales, leading to higher rates of extinction that leave rock specialists isolated phylogenetically.

The second pattern of extensive diversification is consistent with the hypothesis of archipelago speciation, in which geographic speciation is spurred on by the naturally fragmented nature of rock outcrops (Palma-Silva et al., 2011). Archipelago speciation on rock outcrops parallels the iconic examples of plant evolution on oceanic and sky islands systems, but has received far less study (Mota et al., 2020). An alternative, but non-mutually exclusive hypothesis that applies to the second pattern of extensive radiation is ecological release. Edaphic endemism onto bare rocky habitats can be a powerful evolutionary precursor to ecological transitions into other, more open stressful environments, such as harsh chemical soils deserts, cold temperate biomes, and deserts (de Casas et al., 2016; Cacho and Strauss, 2014; Lichter-Marck and Baldwin 2022b). Ecological release predicts that free space from competition opens the door to morphological differentiation and niche divergence (Brown and Wilson, 1956). In the rock specialists, innovations in stress tolerance are predicted to provide the mechanism for shifts into new stressful environments with less competition, and more opportunity for rapid radiation.

**IMPLICATIONS FOR CONSERVATION**

The endemic biodiversity of bare, rocky outcrops and cliffs is threatened by mining, nitrogen deposition, fire, poaching, invasive species, habitat destruction, and climate change (Fitzsimons and Michael, 2017; Corlett and Tomlinson, 2020). The largest extent of rock outcrops however, is found in vertical cliffs in mountainous terrain or deeply eroded canyons, where they are generally inaccessible to people and livestock and therefore free from disturbance (Larson et al., 2000; Kruckeberg, 2004; Fitzsimons and Michael, 2017; Corlett and Tomlinson, 2020; Fig.1). Thus, in landscapes severely impacted by poor management and invasive organisms, cliffs can serve as important refuges for native plants driven to the literal and figurative edge of extinction (Larson et al., 2000; Fitzsimons and Michael, 2017; Batori et al., 2019; La Vigne et al., 2022). On the Hawaiian Islands for example, a hotspot for plant endangerment, cliffs harbor many of the last remaining populations of once widespread plant species driven towards extinction by invasive organisms (Rønsted et al., 2022; La Vigne et al., 2022). Technological innovations in the area of technical safety equipment for repelling down cliffs have enabled botanists to document, collect, and protect endangered plants in previously inaccessible places (Wagner et al, 1994; Lorence and Wagner, 1995; Liittschwager and Middleton, 2001). Technical climbing gear has also spurred a large recreational industry in many places, and this has led to greater access of cliff faces that has caused habitat degradation (Vogler et al., 2011; Fitzsimons and Michael, 2017; Boggess et al., 2021). Efforts to educate climbers about the importance of cliff organisms have generally been successful in mitigating impacts, but we need to do more outreach about the unique biodiversity found on cliffs (Jodice et al., 1999; Fitzsimons and Michael, 2017). The increasing availability of unpersoned aerial vehicles (drones) to researchers has also opened inroads for the discovery of new populations of rare plants, new species, and for systematic plant surveys on previously inaccessible cliffs (Zhou et al., 2021; Nyberg 2019). Recently, La Vigne et al. (2022) introduced a mechanical attachment for remote collection of herbarium specimens and seeds for ex-situ conservation of critically endangered plants from cliffs on Kauai. The prospects for improved exploration and study of cliff plants are exciting both for the prospects of learning about their evolutionary biology and conservation. However, increased accessibility to cliffs comes with risks of habitat degradation, disturbance, and noise pollution (Boggess et al., 2021).

**CONCLUSION**

Bare rock outcrops in cliffs and canyons are striking landscape features revered as iconic landmarks, but our understanding of the specialized plant species found in these habitats remains fragmented. This synthesis represents a first step towards a framework for understanding the processes that cause plant species to exclusively grow on bare rock as well as the contrasting patterns of diversification following edaphic specialization. Much of the accumulated research and insights into plant evolution on other harsh substrates has relevance to understanding adaptation to bare rock. Bare, rocky environments stand apart from other edaphic substrates however, because they have generated stress tolerant lineages that subsequently colonize and diversify in other harsh environments, such as chemically harsh soils and extreme climates. Improved documentation of the heightened plant diversity and endemism on rock outcrops in warmer parts of the world has recently inspired macroecological theories to explain broad-scale biodiversity patterns across space and time (Hopper, 2009; Mucina and Wardel-Johnson, 2011). Likewise, pre-adaptation on rock outcrops can potentially illuminate big questions in plant evolution, such as the 'abominable mystery' of how many distinct lineages of Angiosperms appeared abruptly in the fossil record during the mid-cretaceous (Axelrod, 1972; Davies et al., 2004). Rock outcrops evidently produce altered forms that can prove useful in novel environments, but these novel forms will leave no fossil record because fossilization is rare in rocky places (Looy et al., 2014; Corlett and Tomlinson, 2020). Therefore, the major lineages of Angiosperms could have stemmed from ancestors that evolved their unique characteristics in rock outcrops but did not fossilize until they came down off the rocks to take advantage of new habitats when opportunity arose. This scenario has relevance to our own era of anthropogenic climate change. Given the revolutionary responses of rock plants to stress, these organisms may end up being important sources of adaptive variation that can alleviate the otherwise limited organismal responses to new levels of heat and aridity caused by climate change (Corlett and Tomlinson, 2020). The success of future efforts to conserve biodiversity may therefore depend on our understanding of how unique adaptations for growth on bare, rocky outcrops and cliffs have evolved.

**Acknowledgements**

The author thanks family and colleagues for support and feedback during the preparation of this manuscript, especially Sophia Winitsky, Wren Marck, Bruce Baldwin, Felipe Zapata, Michael Landis, Michael Grundler, Ioana Anghel, Carl Rothfels, Mike R. May, Jenna Bauman, Carrie Tribble, Lukas Mekis, Ben Nyberg, Nina Ronsted, and Kenneth Wood, the Rothfels and Zapata lab groups, the International Compositae Alliance, and the members of the AJB Synthesis committee. The author acknowledges support from NSF-DBI-2209393 and NSF-DEB-2040081. The author declares no conflicts of interest.

Cacho, N.I., A.M. Burrell, A.E. Pepper, and S.Y. Strauss. 2014. Novel nuclear markers inform the systematics and the evolution of serpentine use in Streptanthus and allies (Thelypodieae, Brassicaceae). *Molecular Phylogenetics and Evolution* 72: 71--81.

Cacho, N.I. and P.J. McIntyre. 2020. The role of enemies in bare and edaphically challenging environments. *In* J. Nuñez-Farfan, P.L. Valverde, [eds.], Evolutionary Ecology of Plant-Herbivore Interaction, 249—267. Springer, New York, New York, U.S.A.

Coley, P.D., J. Lokvam, K. Rudolph, K. Bromberg, T.E. Sackett, L. Wright, T. Brenes-Arguedas, D. Dvorett, S. Ring, A. Clark, and C. Baptiste. 2005. Divergent defensive strategies of young leaves in two species of Inga. *Ecology* 86(10): 2633--2643.

Corlett, R.T. and K.W. Tomlinson. 2020. Climate change and edaphic specialists: irresistible force meets immovable object? *Trends in Ecology & Evolution*. 35(4): 367--376.

Davies, T.J., T.G. Barraclough, M.W. Chase, P.S. Soltis, D.E. Soltis, and V. Savolainen. 2004. Darwin's abominable mystery: insights from a supertree of the angiosperms. *Proceedings of the National Academy of Sciences* 101(7): 1904--1909.
20

characterizes the endemic flora of ancient Neotropical mountains. *Proceedings of the Royal Society B*, 287(1923): 20192933.

Zhou, H., J. Zhu, J. Li, Y. Xu, Q. Li, E. Yan, S. Zhao, Y. Xiong, and D. Mo. 2021. Opening a new era of investigating unreachable cliff flora using smart UAVs. *Remote Sensing in Ecology and Conservation* 7(4): 638--648.

**Tables**

**Table 1:** Glossary to technical language used in this synthesis and terms used interchangeably to describe rock dwelling plants and their environments.

| Term | Definition |
| --- | --- |
| Azonal soils | Soils characterized by the influence of local unique soil texture or chemical properties. |
| Calcicole | A plant species often limited to calcareous substrates |
| Chasmophyte | A plant that grows in rock crevices |
| Cliff plants | A plant that grows in vertical exposures of bare rock |
| Edaphic | Pertaining to soil |
| Gypsophile | Plant species tolerant to or restricted to gypsum outcrops |
| Inselbergs | Insular rock outcrops that are surrounded by a matrix of forest, savannah, or desert. |
| Karst | Terrain uniquely sculpted due to high concentrations of calcium carbonate |
| OCBILS | Old Climatically buffered insular landscapes. The concept of OCBILs is based on a hypothesized correlation between high endemism and nutrient poor, old, and climatically stable areas, which often include rock ouctrops and cliffs. |
| Petrophile | A descriptor for plants that prefer rocky environments |
| Rock-dwelling | A plant that facultative grows on rock outcrops |
| Rupicolous | A plant that grows in association with rocks, gravels, scree, or talus |
| Saxicolous | A descriptor for plants that show preference for growth on rocks |
| Zonal soils | Soils or substrates not predominantly under the influence of properties of local unique soil texture or parent material |

**Table 2.** Putative adaptations for growth on bare rock found in rock endemic plant species.

| Trait | Putative benefit | Example | Reference |
|---|---|---|---|
| Dense indument (hairs, trichomes, etc.) | Reduced evapotranspiration, reflectance of UV light, buffer against temperature swings | | Larson et al., 2000 |
| Sclerophyllous (small, waxy) leaves or aphylly (lacking leaves) | Reduced evapotranspiration | | Axelrod, 1967. |
| Active defensive chemistry | Heightened defense against herbivores, potential resilience to cavitation under drought stress | Heightened glucosinolate diversity in jewelflowers (*Streptanthus*) in bare environments | Cacho et al., 2014. |
| Perennial caudex | Anchorage in cracks and crevices, seasonal storage during winter or dry season, regeneration to lessen the burden of recruitment. | Evolution of the perennial caudex in the rock daisies is closely correlated with transitions to bare rock. | Lichter-Marck and Baldwin 2022b. |
| Velloziod roots | Active nutrient uptake by carboxylate release | Velloziaceae species in Campos rupestres in Brazil. | Teodoro et al., 2019 |

| Self-sowing behaviors | Improved establishment of seeds as they are bent back into the rock face | *Eucnide aurea* (A.Gray) H.J.Thomps. & W.R.Ernst and *Laphamia lobata* (I.M.Johnst. ) I.H. Lichter-Marck in Baja California, Mexico. | Rebman et al., 2016 |
|---|---|---|---|
| Reduced seed dispersal elements | Reduced mortality off rock, increased recruitment within rock face | Loss of dispersal elements in blazingstars on edaphic islands of gypsum | Schenk, 2013 |
| Reproductive shifts to self compatibility | Insurance of self-pollination/seed set | Mediterranean *Sonchus* L. (Asteraceae) | Silva et al., 2016 |
| Induction of heat shock proteins or phenolic compounds in leaves | Resistance to high heat and UV exposure | Adaptive plasticity in *Impatiens capensis* Meerb, (Balsaminaceae) grown under high stress conditions | Dixon et al., 2001 |
| Reduced stem elongation | More compact plants experience less exposure to fluctuations in temperature, wind, or desiccation. | Prevalence of ceaspitose growth forms on rock outcrops | Weinig et al., 2004. |

**Table 3.** A selected list of lineages with abundant rock-endemic species from western North America and a qualitative assessment of patterns of diversification (see Box 2 for illustrations of Pattern I and II).

| Lineage (Family) | Number of species | Examples of rock-dwelling taxa | Pattern of lineage diversification | Reference |
|---|---|---|---|---|
| Ivesioids (Rosaceae) | ~50 spp. In three genera *Ivesia* Torr. & A.Gray, *Horkelia* Cham. & Schltdl., and *Horkeliella* (Rydb.) Rydb. | *Ivesia baileyi* S.Watson, *Ivesia arizonica* (J.T.Howell) Ertter | Pattern I | Topel et al., 2012. |
| *Agave* L. (Asparagaceae) | >250 spp. | *Agave petrophila* García-Mend. & E.Martínez, *Agave Ocahui* Gentry, *Agave victoriae-regia* T.Moore | Pattern I | Jiménez-Barron et al., 2020. |
| *Penstemon* Schmidel (Plantaginaceae) | ca. 285 | *Penstemon baccharifolius* Hook.*,* *Penstemon calcareus* Brandegee*,* *Penstemon caryi* Pennell | Pattern I | Wolfe et al., 2021. |
| *Ericameria* Nutt. (Asteraceae) | Ca. 36 | *Ericameria cuneata* McClatchie*,* *Ericameria obovata* (Rydb.) G.L.Nesom | Pattern I | Roberts and Urbatsch, 2003 |

| | | | | Noyes, 2000. |
|---|---|---|---|---|
| *Erigeron* L. (Asteraceae) | Ca. 390 | *Erigeron salmonensis* Brunsfeld & G.L.Nesom*, E. tener* M.E. jones*, E. rupicola* Phil. (Chile) | Pattern I | |
| *Tribe Spiraeae* Nutt. (Rosaceae) | Ca. 120-180 | Genus *Petrophytum* (Nutt.) Rydb. (3 spp), *Kelseya uniflora* (S.Watson) Rydb. (monotypic) | Pattern I | Potter et al., 2007 |
| *Diplacus* Nutt. + *Erythranthe* Spach (Phrymaceae) | Ca. 170 | *Diplacus rupicola, Erythranthe eastwoodieae* | Pattern I | Beardsley et al., 2004. |
| Saxifragaceae s.s. | ~600 spp. mostly in the genus saxifraga | *Telesonix jamesii* Raf. (monotypic), *Heuchera* L. (ca. 37 spp) | Pattern II | De Casas et al., 2016; Folk et al., 2021. |
| *Eucnide* Zucc. (Loasaceae) | Ca. 15, most restricted to rock outcrops and cliffs | *Eucnide aurea, Eucnide urens* Parry | Pattern II | Brindley et al., 2021 |
| *Boechera* Á.Löve & D.Löve (Brassicaceae) | ca. 110 | *Boechera depauperata* (A.Nelson & P.B.Kenn.) Windham & Al-Shehbaz | Pattern II | Alexander et al. 2013 |
| Perityleae (Asteraceae) | Ca. 57 | *Laphamia cochisensis* W.E. Niles, *Laphamia inyoensis* Ferris, *Laphamia cernua* Greene | Pattern II | Lichter Marck et al. 2020; Lichter-Marck and Baldwin, 2022b |
| Cheilenthoideae (Pteridaceae) | *Myriopteris* Fee *(47)*, *Astrolepis* D.M.Benham & Windham *(6)*, *Pellaea* Link *(35)*, *Bommeria* E.Fourn. *(5)*, *Argyrochosma* (J.Sm.) Windham *(20)*, *Aspidotis* (Nutt. ex Hook.) Copel. *(4)*, *Pentagramma* Yatsk., Windham & E.Wollenw. *(6)*, *Notholaena* R.Br.*(34)* | *Argyrochosma jonesii* (Maxon) Windham, *Myriopteris lindheimeri* J.Sm. | Pattern II | Scheuttpelz et al., 2007 |
| *Mammillaria* Haw. s.l. (Cactaceae) | Ca. 164-200 | *Mammillaria grahamii* Engelm. | Pattern II | Breslin et al., 2021 |
| *Maurandyinae* (Plantaginaceae) | *Mabrya* Elisens, *Holmgrenanthe* Elisens | *Holmgrenanthe petrophila* (Coville & C.V.Morton) Elisens, *Mabrya acerifolia* (Pennell) Elisens | Pattern II | Ogutcen and Vamosi, 2016 |

**Table 4:** Outstanding questions.

| |
|---|
| Is the evolution of edaphic specialization onto bare rock clustered across the phylogeny of plants? |
| What role does polyploidy and hybridization play in providing the raw material for adaptation to bare rock faces in plants? |
| What determines whether a lineage specialized onto rock outcrops remains evolutionarily isolated (pattern I) or undergoes an extensive radiation (pattern II)? |
| What is the relative importance of rockiness in determining habitat bareness compared to other factors, such as chemistry, climate, or natural enemies? |
| How can conservation approaches be altered to better serve rare plants on cliffs and bare rocky outcrops? |
| What are the genomic underpinnings of adaptation to bare rock in plants? |

**Figures**

**Figure 1:** The largest extent of rock outcrops is found in vertical cliffs in mountainous terrain or deeply eroded canyons. (A) Rhyolite cliffs in cave creek, Chiricahua mountains, Cochise County, Arizona, U.S.A. (B) Deeply eroded cliffs of Kalalau valley on the leeward side of Kauai, Hawaii, U.S.A. (C) Patchy rock outcrops in otherwise densely vegetated tropical deciduous forests in the western Sierra Madre Occidental in Sonora, Mexico. (D) Outcrop of limestone in Great Basin Desert on Conglomerate mesa, Inyo County, California. Photos by ILM.

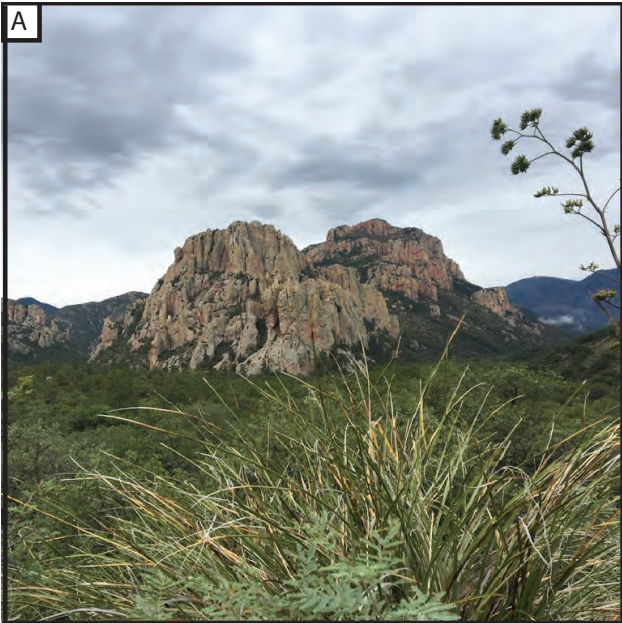
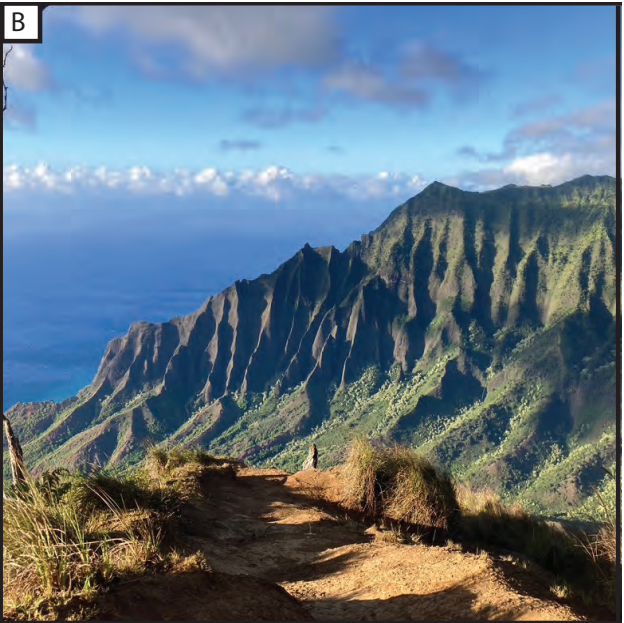
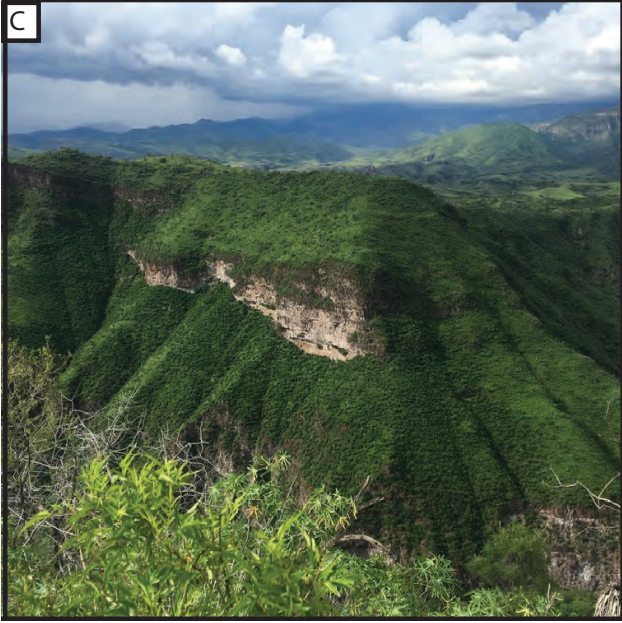
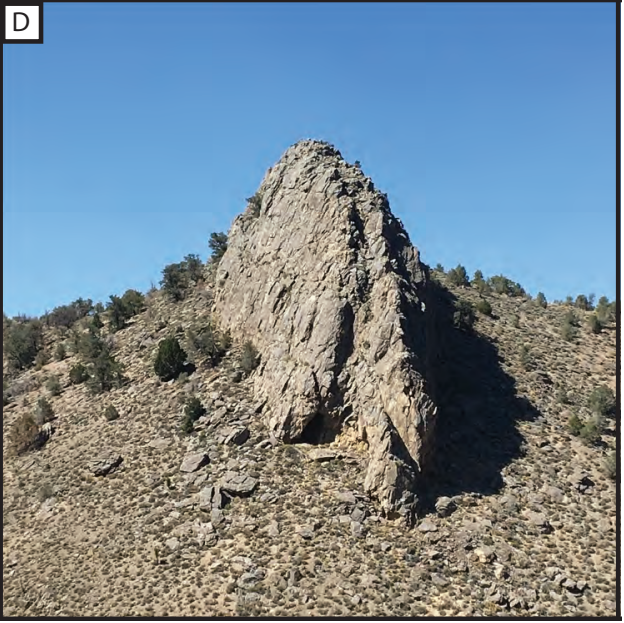

**Figure 2:** A diverse assemblage of rock endemic plant taxa that are found exclusively on cliffs in the canyons and mountains of Death Valley National Park in California, U.S.A. (A) *Diplacus rupicola* (Phrymaceae). (B) *Holmgrenanthe petrophila* (Plantaginaceae). (C) *Phacelia perityloides* Coville (Boraginaceae). (D) *Argyrochosma jonesii* (Pteridaceae). (E) *Hesperidanthus jaegeri* (Rollins) Al-Shehbaz (Brassicaceae). (F) *Dudleya saxosa* (M.E.Jones) Britton & Rose (Crassulaceae). (G) *Petrophytum caespitosum* (Rosaceae). (H) *Laphamia* cf. *villosa* S.F. Blake (Asteraceae). Photos:

A:David Greenburger; B:Teague Embrey; C,E: Steve Matsen; D,E,G,H:ILM.

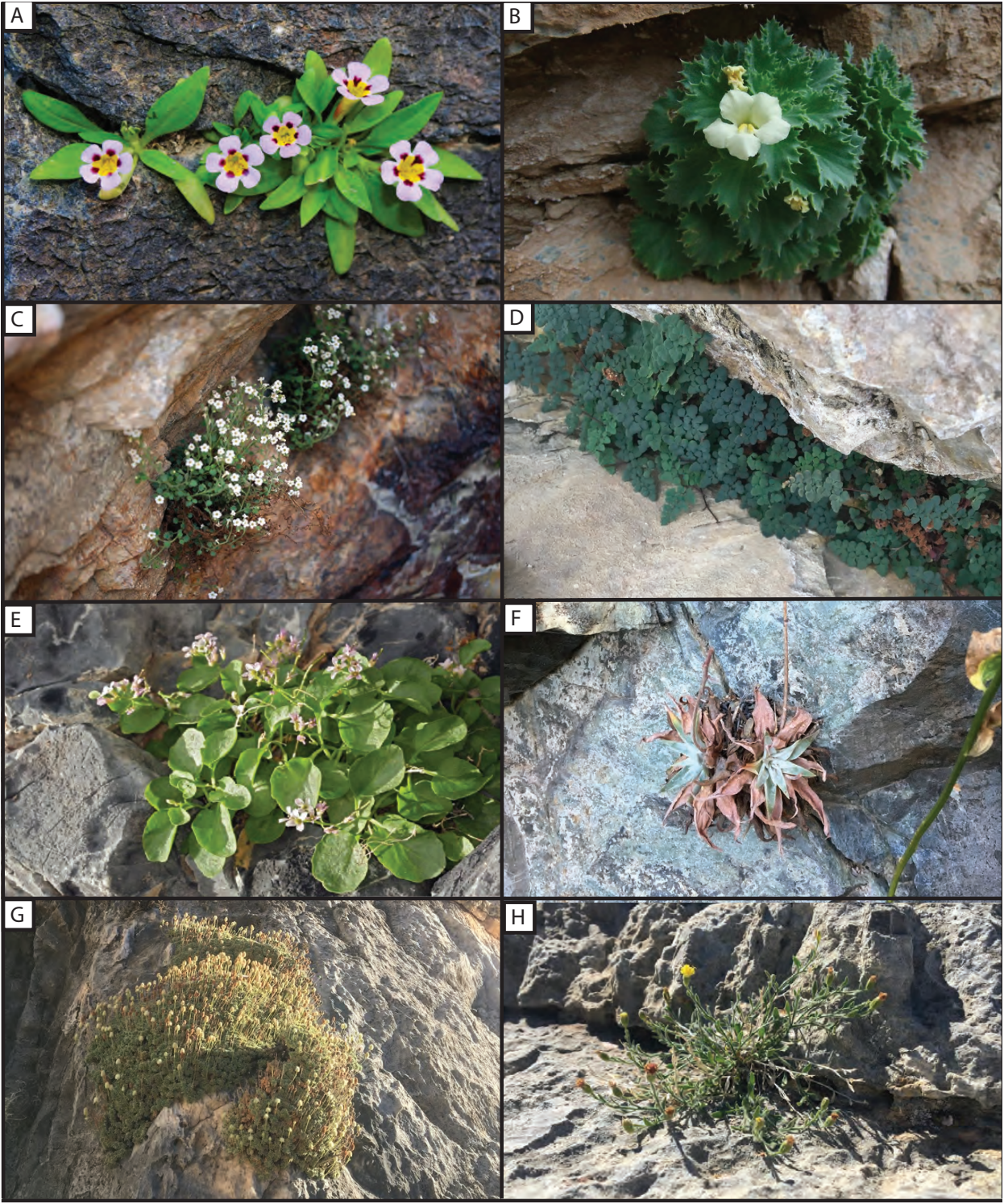

**Figure 3:** Two species of rock-endemic plants that are found together on cliffs in Baja California have independently evolved a specialized behavior for reduced dispersal. During fruit maturation the peduncle bends back towards the rock face, reseeding itself in cracks below the parental plant. (A) *Eucnide aurea*, flower. (B) Eucnide aurea, habitat on volcanic rock face in Sierra de la Giganta, Baja California Sur, Mexico. (C) Mature infructescence of *Eucnide aurea* bending back to self-sow in a crevice. (D) *Laphamia lobata*, habitat on volcanic rock cliffs in San Basilio, Baja California Sur, Mexico. (E) Mature heads bending back towards the cliff. (F) Similar behavior demonstrated by *Laphamia leptoglossa*, a related rock daisy found on volcanic cliffs in the Madrean sky islands of Sonora, Mexico.

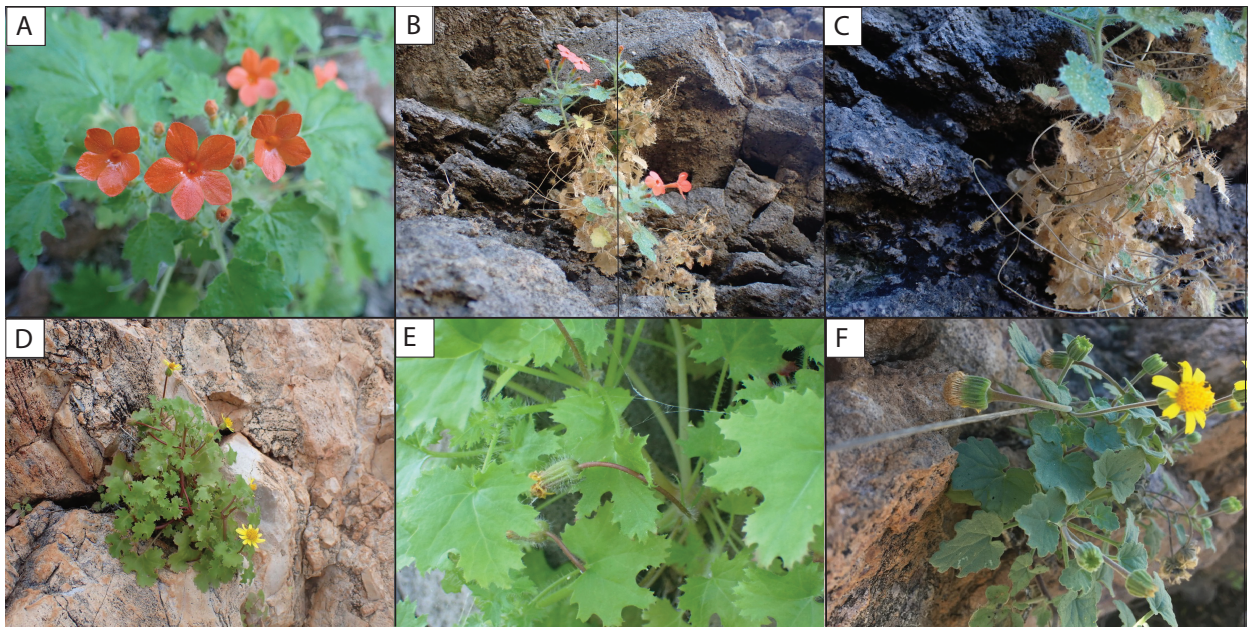

**Figure 4:** Genus *Laphamia* of the rock daisy tribe (Perityleae), an exemplary case of plant diversification on rock outcrops, comprises 57 species of rock specialists that have undergone an extensive radiation in habitat islands, including edaphic islands, throughout the Great Basin province of western North America. (A) *Laphamia leptoglossa* (Harv. & Gray ex A.Gray) I.H.Lichter-Marck. (B) *Laphamia villosa*. (C) *Laphamia cernua*. (E) *Laphamia quinqueflora* Steyerm.. (F) *Laphamia cochisensis*. (G) *Laphamia rupestris* var. *albiflora* (A.M.Powell) I.H.Lichter-Marck. (H) Ancestral state reconstruction of obligate growth on bare rock in the rock daisy tribe (Perityleae) shows that edaphic specialization began early and is a shared derived characteristic of most extant species. These results are derived from reverse jump MCMC analysis in RevBayes. Orange circles correspond to growth on decomposed substrates and blue circles to bare rock. The size of the circles corresponds to their posterior probability. Circumference colors correspond to generic limits following recent reclassification. More details on analyses and taxonomic limits can be found in Lichter-Marck & Baldwin (2022a, 2022b). Photos A-G: ILM. H is modified from Lichter-Marck and Baldwin (2022b) with permission of the authors.

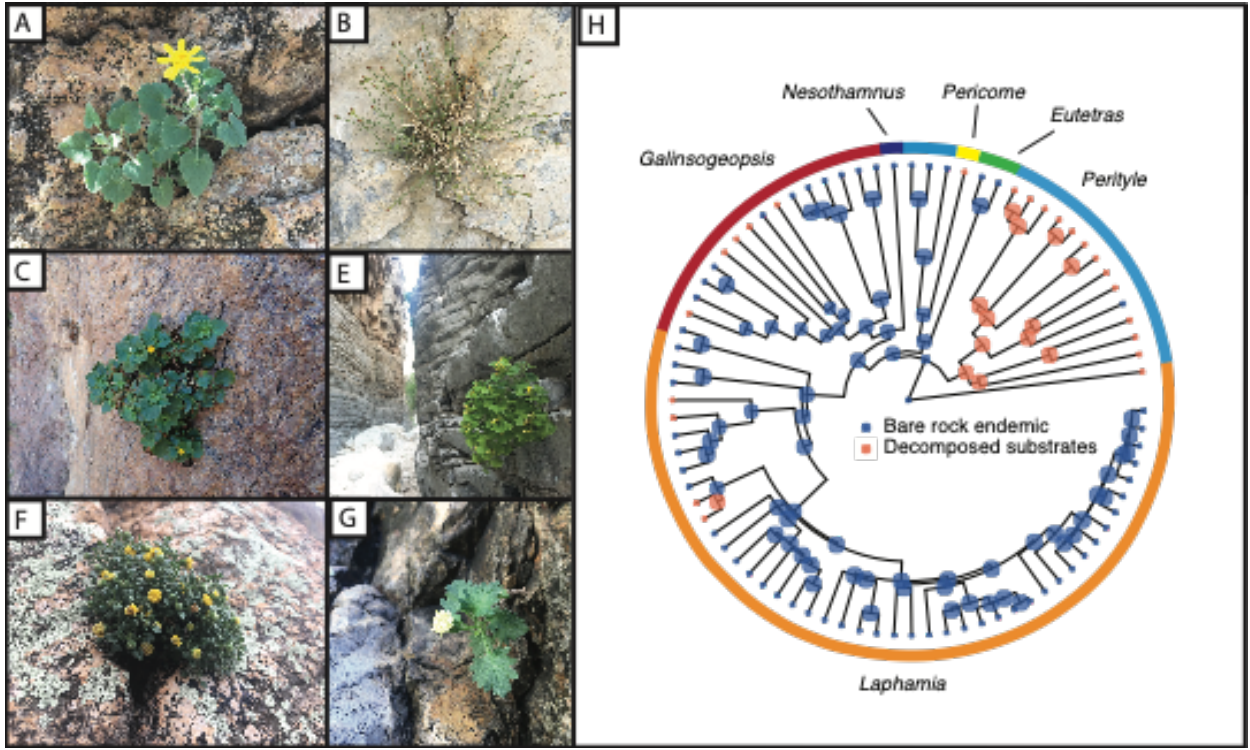

**Box 1:** What are the evolutionary processes that lead to edaphic specialization onto bare rocky outcrops?

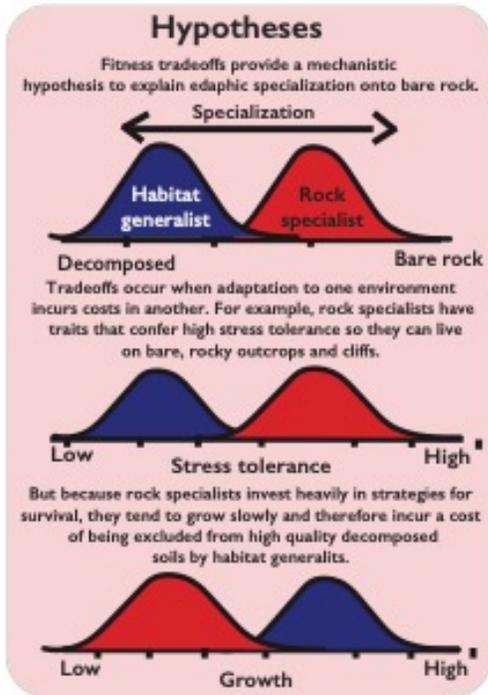

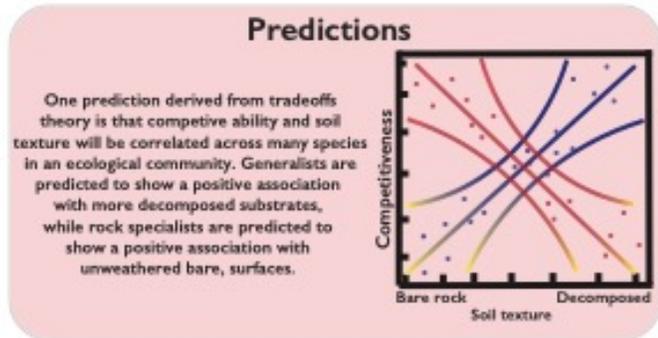

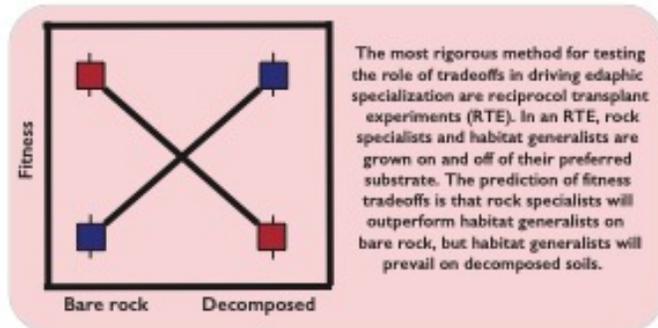

**Box.2:** What are the consequences of edaphic specialization on lineage diversification?

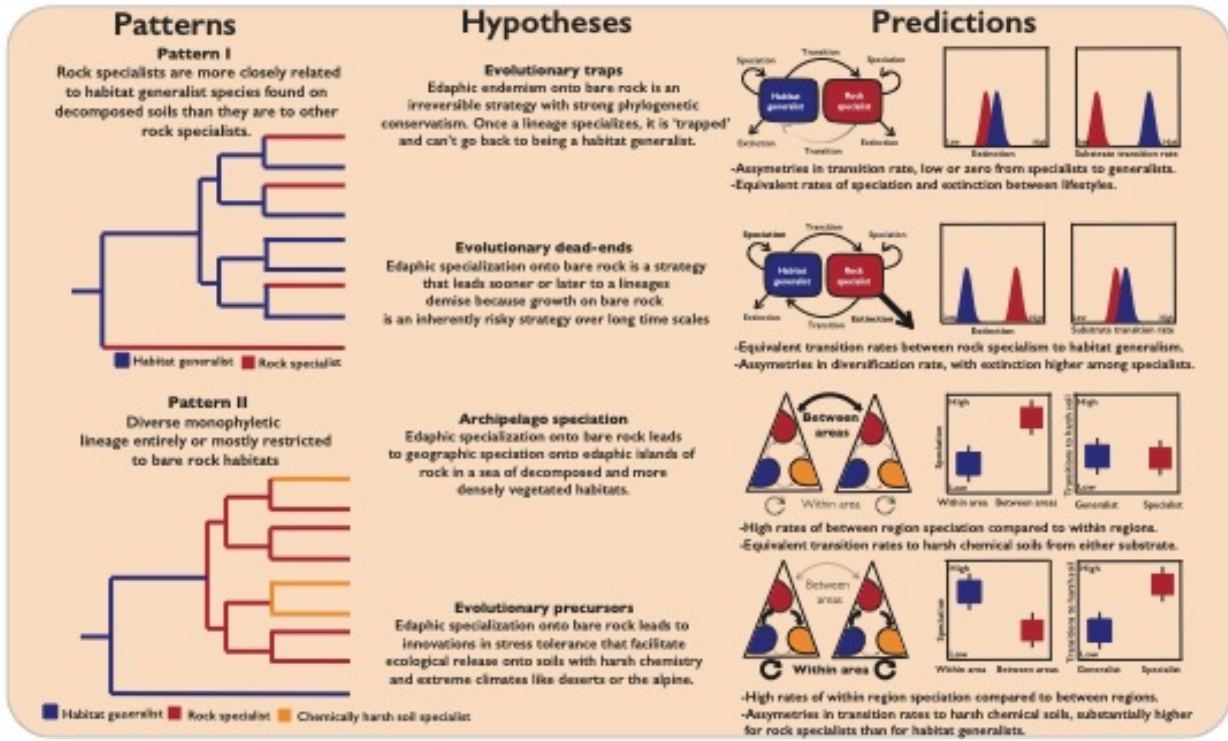